\documentclass[aps,pre,twocolumn,superscriptaddress]{revtex4-1}
\usepackage[dvipdfm]{graphicx}
\usepackage{amssymb}
\usepackage{here}

\begin{document}


\title{Collective motion of oscillatory walkers}


\author{Takahiro Ezaki}
\email{ezaki@jamology.rcast.u-tokyo.ac.jp}
\affiliation{Department of Aeronautics and Astronautics, Graduate School of Engineering, The University of Tokyo, 7-3-1 Hongo, Bunkyo-ku, Tokyo 113-8656, Japan}
\affiliation{Japan Society for the Promotion of Science, 8 Ichibancho, Kojimachi, Chiyoda-ku, Tokyo 102-8472, Japan}
\author{Ryosuke Nishi}
\altaffiliation[Present address: ]{Department of Mechanical and Aerospace Engineering, Graduate School of Engineering, Tottori University, 4-101 Minami, Koyama, Tottori 680-8552, Japan}
\affiliation{
National Institute of Informatics, 2-1-2 Hitotsubashi, Chiyoda-ku, Tokyo 101-8430, Japan
}
\affiliation{
JST, ERATO, Kawarabayashi Large Graph Project, 2-1-2 Hitotsubashi, Chiyoda-ku, Tokyo 101-8430, Japan
}

\author{Daichi Yanagisawa}
\altaffiliation[Present address: ]{Department of Aeronautics and Astronautics, Graduate School of Engineering, The University of Tokyo, 7-3-1 Hongo, Bunkyo-ku, Tokyo 113-8656, Japan}
\affiliation{College of Science, Ibaraki University,  2-1-1, Bunkyo, Mito, Ibaraki, 310-8512, Japan}

\author{Katsuhiro Nishinari}
\affiliation{Research Center for Advanced Science and Technology, The University of Tokyo, 4-6-1, Komaba, Meguro-ku, Tokyo 153-8904, Japan}


\date{\today}

\begin{abstract}
We study a system of interacting self-propelled particles whose walking velocity depends on the stage of the locomotion cycle.
The model introduces a phase equation in the optimal velocity model for vehicular traffic. 
We find that the system exhibits novel types of flow: synchronized free flow, phase-anchoring free flow, orderly jam flow, and disordered jam flow.
The first two flows are characterized by synchronization of the phase, while the others do not have the global synchronization.
Among these, the disordered jam flow is very complex, although the underlying model is simple.
This phenomenon implies that the crowd behavior of moving particles can be destabilized by coupling their velocity to the phase of their motion.
We also focus on ``phase-anchoring" phenomena. They strongly affect particle flow in the system, especially when the density of particles is high.
\end{abstract}

\pacs{}

\maketitle

\section{introduction}
The movements of all animate things, such as walking of humans, flapping of birds, and swimming of fish,
are restricted by their locomotion \cite{loco}.
Since the collective motion of self-propelled particles (SPP) began to attract interest \cite{vic,vic2,vic3},
many studies have been conducted in various contexts \cite{etc}, including pedestrian crowds \cite{ped,ped2,ped22,ped3,Prox,ped4,ped5,Exp,Rhythm}, bird flocks \cite{flock}, and insect swarms \cite{ins,ins2}.
However, the effect of locomotion on macroscopic behavior remains an open question.
In fact, recent experiments in the field of pedestrian dynamics suggest that the oscillatory motion
 of pedestrian walking may have a significant effect on overall dynamics \cite{Exp,Rhythm}. 
Jeli\'c \textit{et al}. studied the collective dynamics of pedestrians walking in a line, focusing on bilateral oscillations during
walking \cite{Exp}. From fluctuating trajectories, they obtained the phase of locomotion and investigated
the interaction between neighboring pedestrians. 
Furthermore, Yanagisawa \textit{et al.} \cite{Rhythm} reported that synchronization of the locomotion phase among pedestrians may
increase the total flux.
Here the ``phase" is a mapping from each stage of locomotion 
to a real number  $\phi \in \mathbb{R}/2\pi\mathbb{Z}$.  Unlike the situation in the vehicular traffic,
this phase is strongly related to the velocity of walkers.
\begin{figure}[H]
   \includegraphics[width=90mm]{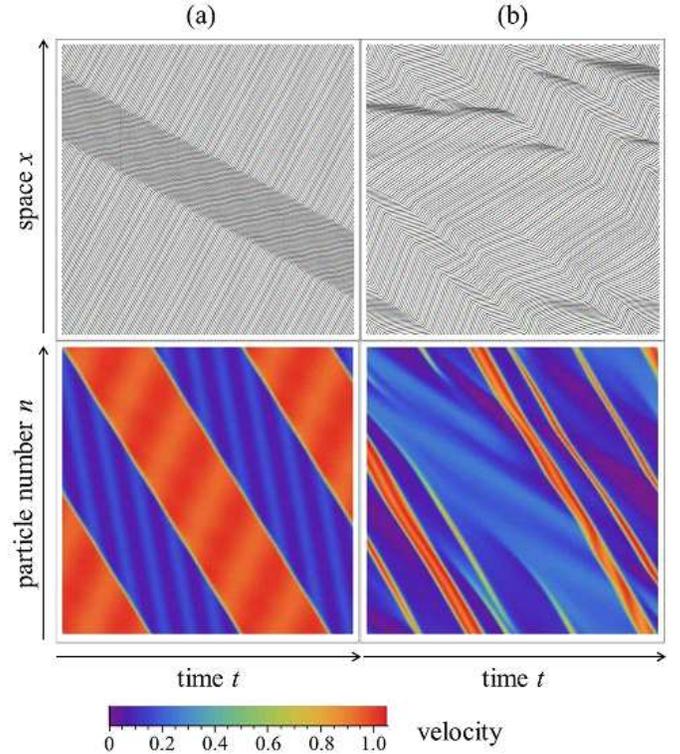}
 \caption{Space-time diagrams (top) and velocity-time diagrams (bottom) for orderly jam flow (left, $\rho=2.0, K=5.0,A=0.05$) and disordered jam flow (right, $\rho=3.0, K=1.0,A=0.05$). 
 Bold lines indicate the trajectory of the 50th walker.
The small regular waves in orderly jam flow correspond to phase anchoring. }
 \label{ts}
\end{figure}

In this paper, we propose a simple model that describes particle-following behavior, phase-velocity coupling, and phase-phase coupling.
Particle-following behavior is implemented by the optimal velocity (OV) model \cite{OV} for vehicular traffic, 
which is known to emerge the phase transition from free flow to jam flow that is highly tractable in mathematical analysis.
Phase-phase coupling is considered within the framework of the Kuramoto model \cite{kuramoto},  a paradigmatic model describing synchronization phenomena in nature.
In the present study we assume that the phases of successive particles tend to synchronize, which is 
experimentally suggested when pedestrians walk in a dense crowd \cite{Exp}.
Since each walker follows its predecessor and ignores its successors, the interaction is unidirectional and local.
In contrast, other studies have been mainly
devoted to globally coupled oscillators or locally but bidirectionally coupled oscillators \cite{kuramotoL,kuramotoRev}. 
Note that we do not consider detailed modeling of actual animals; instead, we concentrate on ideal particles called oscillatory walkers (OWs).
Although here we assume the interaction between particles is through their headway distances, it could depend also on their (relative) velocities, 
which is beyond scope of this paper.

In spite of its simplicity, the OW model presents a rich behavior, including a novel type of jam [Fig. \ref{ts}(b)]. 
We report its fundamental characteristic and give insight into collective locomotion.

The rest of the paper is organized as follows: The next section is devoted to define the model. In Sec. \ref{dyn} we summarize the dynamics of the model.
To understand its destabilization phenomena, theoretical analyses are performed in Sec. \ref{theo}. 
Finally we summarize these results in the final section. 

\section{model}
Consider $N$ particles, labeled $n=1,\cdots,N$ from left, walking to right 
on a one-dimensional ring. We impose periodic boundary conditions 
[particle $(N+1)$ particle is identical to particle $1$].
The velocity of the $n$th walker ($\dot{x}_n$) is determined by the following equation:
\begin{equation}
\ddot{x}_{n} = a\{ V(\Delta x_{n}) + A (\cos{\phi_n}+1) - \dot{x}_n\}\label{se1},
\end{equation}
where $V(\Delta x_n)$ is the optimal velocity determined for headway distance ($\Delta x_n=x_{n+1}-x_n$)  \cite{OV}.
Each walker adapts its velocity to this optimal velocity with an adaptation intensity (the reciprocal of the reaction time) $a$. 
The effect of locomotion is represented by the oscillation term $A(\cos{\phi_n}+1)$. 
The quantity $A$ is the magnitude of the fluctuations in the target velocity, which is set to a small value.  
The fluctuation phase $\phi_n$ is 
determined by 
\begin{equation}
\dot{\phi}_{n} = \omega(\Delta x_{n}) + K \sin{\Delta\phi_{n}}.\label{se2}
\end{equation}
Here we assume that the ideal angular velocity $\omega (\Delta x_n)$ is defined by a function similar to
 the ideal walking velocity $V(\Delta x_n)$, i.e., using a normalized optimal velocity (NOV) function $U(\Delta x_n)$,
 we can express $\omega(\Delta x_n)=\Omega_MU(\Delta x_n)$ and $ V(\Delta x_n)=V_MU(\Delta x_n)$ .
We further impose the condition that when the headway distance is large enough, 
particles walk with a steady angular velocity $\Omega_MU_{\infty}$ and velocity $V_MU_{\infty}$, 
and for small $\Delta x_n$, $U(\Delta x_n)\sim 0$ to stop.
To capture these assumptions, the NOV function is set to be a differentiable and monotonically increasing function that is asymptotic to $0$ as 
$\Delta x_n\rightarrow 0$ and to a positive constant as $\Delta x_n\rightarrow \infty$. 
We use the following form of the NOV function when actual calculations are needed:
$U(\Delta x_n) = c_1 [ \tanh (c_2 \Delta x_n - c_3) + \tanh c_3 ]$ with the positive parameters $c_1, c_2$ and $c_3$. $U(\Delta x_n)$ is a monotonic increasing function such that 
$U(0) = 0$ and $U(\infty) = c_1 [ 1 + \tanh c_3 ]$. The quantity $c_3/c_2$ corresponds to its inflection point, 
and $c_1$ and $c_3$ determines its height given by $c_1 [ 1 + \tanh c_3 ]$. 
Here we set $c_1 = 0.5, c_2 = 5$, and $c_3 = 2.5$.

The second term on the rhs of Eq. (\ref{se2}) represents the synchronous interaction 
of successive walkers, whose intensity is $K>0$. 
When the phase difference $\Delta \phi_n=\phi_{n+1}-\phi_{n}$ is zero and the walkers are uniformly distributed,
they walk with a steady velocity and rhythm (synchronized free flow). 
In this paper, we restrict the model parameters 
to $a=3,V_M=1, \Omega_M=1$ and $N=100$. We vary the particle density $\rho = N/L$
by changing the system length $L$.
\begin{figure*}[t]
\includegraphics[width=170mm]{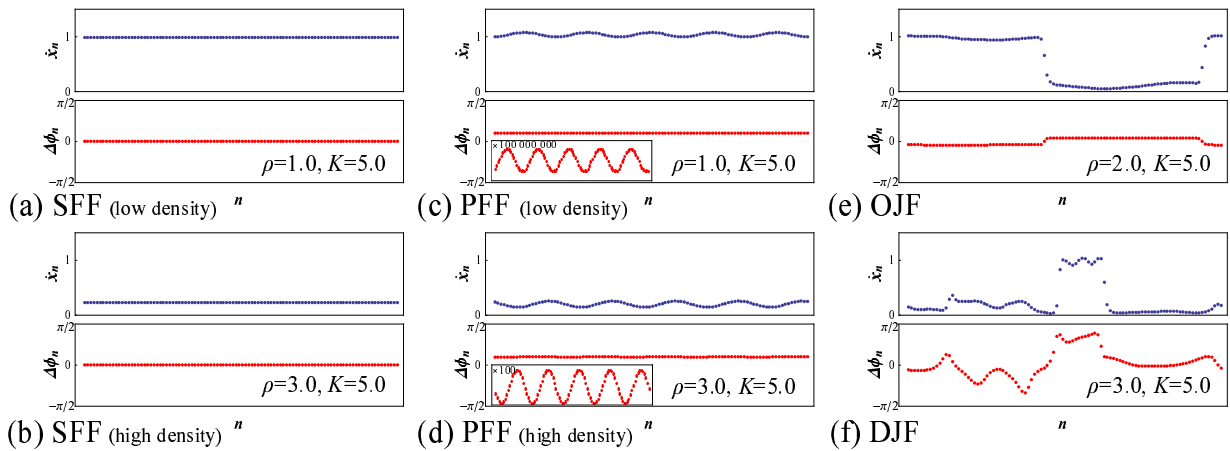}
\caption{Snapshots of velocity and phase difference for $A=0.05$.
On the abscissae, $n$ represents particle number.
Panels (a) and (b) are for synchronized free flow (SFF), panels (c) and (d) are for phase-anchoring free flow (PFF), panel (e) is for orderly jam flow (OJF), and panel (f) is for disordered jam flow (DJF). 
Wavy shapes of velocity in (c), (d), and (e) stem from the constant phase difference. 
In PFF (c) and (d), the phase differences are almost constant, but slight fluctuations occur, as shown in the insets. For $a=3.0$ we set, $\rho=1.0,2.0$ and $3.0$ correspond to low-, middle-, and high-density regimes, respectively. As explained in the text, the middle-density regime has OJF only.}
\label{fig:two}
\end{figure*}

\section{Dynamics of the OW model}\label{dyn}
The OW model has four types of flow: synchronized free flow (SFF), phase-anchoring free flow (PFF), orderly jam flow (OJF), and disordered jam flow (DJF). 
``Phase anchoring" occurs when phase differences $\Delta\phi_n$ are fixed at a 
common value, $\Delta \phi_0\in (-\pi,\pi]$. When the effects of distance between walkers are negligible, 
the model reduces to the Kuramoto model with local interactions. Then, the states $\Delta\phi_n=$const.,
in which all angular velocities are identical, are stable over a certain range of $\Delta\phi_0$. 
This property is inherited by the OW model; however, as can be easily verified, the states 
$\Delta \phi_n=$const.$\neq 0$ cannot be realized exactly because of the effects of spatial structure.
Actually, each phase difference fluctuates slightly around $\Delta\phi_0$ as depicted in Figs. \ref{fig:two}(c) and \ref{fig:two}(d).

Before going into details, we briefly summarize the dynamics in the system. 
First, similar to vehicular traffic flow, the model has phase transitions between free flow and (orderly) jam flow; these are caused by increases in the density of walkers.
When the density is small ($\rho<\rho_{\rm{cr,1}}$), free flow is stable and a jam cannot develop.  
At middle densities $\rho_{\rm{cr,1}}<\rho<\rho_{\rm{cr,2}}$, free flow is no longer stable, and small perturbations 
grow into  a jam; this is called OJF [Figs. \ref{ts}(a) and \ref{fig:two}(e)]. 
At high densities $\rho_{\rm{cr,2}}<\rho$, the system regains stability for free flow. 
In the free flow regime, the OW model has two types of flow: SFF [Figs. \ref{fig:two}(a) and \ref{fig:two}(b)] and PFF [Figs. \ref{fig:two}(c) and \ref{fig:two}(d)].
These flows are locally stable, and all transient flows are asymptotic to either SFF or PFF in the low-density regime, depending on the initial condition.
In contrast, in the high-density regime, SFF is always locally stable, while
 PFF can be destabilized for some parameter regimes, resulting in another type of jam.

In OJF of the middle-density regime, separation of phase differences occurs.
As shown in Fig. \ref{fig:two}(e), the phase differences become almost constant in each region. 
This phase anchoring is produced by the spatial structure of OJF, and it also depends on $K$ and the 
initial conditions.

In addition to these states, another nontrivial phenomenon emerges in the high-density regime, namely disordered jam flow (DJF).
This is complex jam flow induced by strong coupling between $\phi_n$ and $x_n$. 
As shown in Figs. \ref{ts}(b) and \ref{fig:two}(f), no steady pattern occurs, and small jams propagate with different velocities 
and even disappear and reappear.
The four types of flow correlate with density as in Table. \ref{table1}. 

Now, we investigate the system in detail.
\begin{table}[t!]
\caption{Classification of flows. Each tick indicates the existence of the flow in the density regime. }\label{table1}
\begin{tabular}{c|c|c|c|c}
\hline\hline
\multicolumn{1}{c}{} & \multicolumn{1}{c|}{} & \multicolumn{3}{c}{Density} \\
\cline{3-5}
\multicolumn{1}{c}{} & \multicolumn{1}{c|}{} & Low & Middle & High \\
\hline
\raisebox{-1.8ex}[0pt][0pt]{Free flow} & Synchronized free flow (SFF)& \checkmark &  & \checkmark \\
\cline{2-5}
 & Phase-anchoring free flow (PFF)& \checkmark &  & \checkmark \\
\hline
\raisebox{-1.8ex}[0pt][0pt]{Jam flow} & Orderly jam flow (OJF)&  & \checkmark &  \\
\cline{2-5}
 & Disordered jam flow (DJF)&  &  & \multicolumn{1}{|c}{\checkmark} \\
\hline\hline
\end{tabular}
\end{table}
\subsection{Phase-anchoring phenomena}
We first focus on phase-anchoring phenomena. 
In this paper, we use the following definition for PFF: 
(i) Each $\Delta \phi_n$ and $\Delta x_n$ periodically fluctuates around a certain phase difference $\Delta\phi_0$ (const.) and headway distance $\Delta x_0 =\frac{1}{\rho}$, respectively. 
(ii) The ranges of these fluctuations are small $(\ll|\Delta\phi_0|)$.

\begin{figure}[t!]
   \includegraphics[width=90mm]{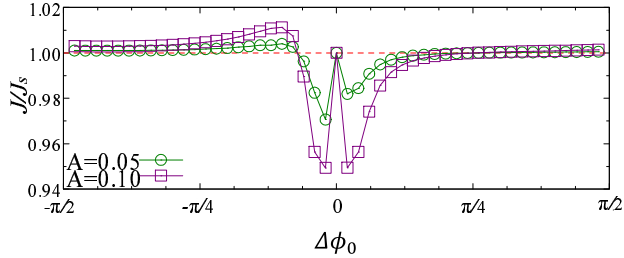}
 \caption{Relationship between flux and phase difference for the case of $\rho=4.0$ and $K=5.0$ (PFF in the high-density regime). 
Here the flux is normalized by the SFF value ($\Delta \phi_0=0$).}
 \label{flowchange}
\end{figure}

The effects of phase anchoring are significant, especially when the density of walkers is high.
In this regime, since each $\Delta x_n$ and $\omega(\Delta x_n)$ is small, the dependence of the system dynamics on phase becomes relatively large.
One intriguing feature is the relationship between phase anchoring and the flux of walkers.
We show the relation between flux and phase difference $\Delta \phi_0$ in Fig. \ref{flowchange}.
Flux is defined as the product of density and average velocity, averaged over time, $J=\langle \rho \bar{\dot{x}}\rangle$. For SFF, we can calculate the flux as 
$J_s = \rho \left[V(\frac{1}{\rho})+A\right]$. Simulation results indicate that for small phase delays and advances,
 the flux is decreased, but for large phase differences, the flux increases, especially when the phase advances ($\Delta\phi_0<0$).
In addition, as expected, the effect of phase difference is conspicuous when the intensity of the coupling parameter $A$ is large.
Although phase-anchoring can be observed in the low-density regime, there is little change in the flux because 
when the headway distance is large, fluctuations in phase anchoring do not influence the angular velocity 
through the term $\omega(\Delta x_n)\simeq \Omega_M$. 
In PFF, each $\Delta x_n$ oscillates with a certain 
phase that differs from the walking phase $\phi_n$. This difference distorts the trajectory of walkers from that in SFF (represented by a sine curve),
leading to decreases and increases in the flux. (See also Appendix \ref{sec:dis}.)

\subsection{Disordered jam flow}
\begin{figure}[t!]
   \includegraphics[width=87mm]{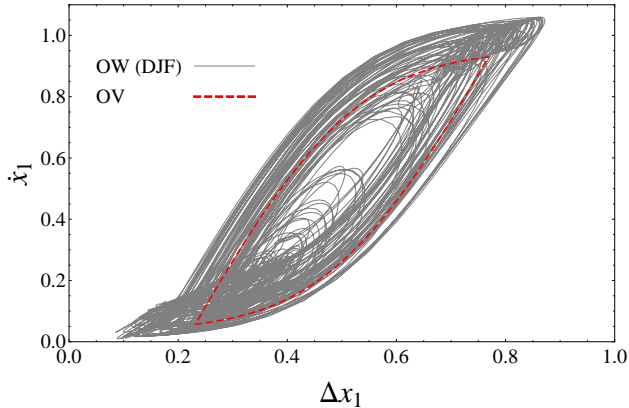}
 \caption{Headway-velocity trajectory for DJF ($A=0.05,\rho=3.0,$ and $K=1.0$). The red broken line represents the trajectory of jam flow in the OV model ($A=0, \rho=2.0$).}
 \label{lim}
\end{figure}

For small $K$ and high density $\rho$, free flow allows only synchrony ($\Delta \phi_0=0$) and PFF becomes unstable,
and a highly complex flow appears. This flow is phenomenologically distinct from OJF,
which is already observed in the OV model. Instead, it originates from the strong coupling between $x_n$ and $\phi_n$.
In DJF, small jams locally appear and disappear intermittently, and no stationary state is reached [Fig. \ref{ts}(b)]. 
Further, we can no longer observe regular patterns in $x_n$, $\dot{x}_n$, or $\Delta \phi_n$ [Fig. \ref{ts}(b) and \ref{fig:two}(f)]. 
Figure \ref{lim} shows a complex limit cycle of DJF in the $(\Delta x_1,\dot{x}_1)$-plane; the figure also shows the trajectory of the jam flow in the OV model. 
In the quasi-periodic trajectory of DJF, we can see distinctive patterns, especially at its edges. 

\section{theoretical analyses}\label{theo}
Next, we discuss the stability conditions for each flow. 
To understand the destabilization of steady free flow, $(x_n^{(0)},\phi^{(0)}_n)$, 
we investigate how the flow responds to small perturbations.
For a small perturbation, $x_n= x_n^{(0)} + \epsilon_n, \phi_n = \phi_n^{(0)} +\delta_n$, the system equations (\ref{se1}) and (\ref{se2}) linearize to
\begin{eqnarray}
\ddot{\epsilon_n}/a &=& - \dot{\epsilon_n} + V_MU'(\Delta x^{(0)}_n)\Delta\epsilon_n - A \sin{\phi_n^{(0)}}\delta_n,\label{epg}\\
\dot{\delta_{n}} &=& K\cos{\Delta \phi^{(0)}_n}\Delta\delta_n+\Omega_MU'(\Delta x^{(0)}_n)\Delta\epsilon_n\label{deg}
\end{eqnarray}
with $U'(\Delta x^{(0)}_n)=\frac{\partial U}{\partial\Delta x}|_{\Delta x=\Delta x^{(0)}_n}$.
When all walkers are uniformly synchronized, namely, $\Delta x_n^{(0)}=L/N=1/\rho$, $\phi_n^{(0)}=\Omega_M U(1/\rho)t\equiv \dot{\phi}_s t$,
Eqs. (\ref{epg}) and (\ref{deg}) simplify to
\begin{eqnarray}
\ddot{\epsilon_n}/a &=& - \dot{\epsilon_n} + V_MU'(1/\rho)\Delta\epsilon_n - A \sin{(\dot{\phi}_st)}\delta_n,\label{ep}\\
\dot{\delta_{n}} &=& K\Delta\delta_n+\Omega_MU'(1/\rho)\Delta\epsilon_n\label{de}.
\end{eqnarray}
Here we assume that $\dot{\phi}_st$ is a fast variable compared to the growth rate of the perturbation,
i.e., $\epsilon$, $\delta$, and their derivatives are constant for each period $T=\frac{2\pi}{\dot{\phi}_s}$.
By integrating Eq. (\ref{ep}) over this period, we can eliminate the vibrational term, leaving
\begin{equation}
\ddot{\epsilon_n}/a = - \dot{\epsilon_n} + V_MU'(1/\rho)\Delta\epsilon_n\label{ep2},
\end{equation}
which is the linearized OV model. Now, the stability condition is given by $V_MU'(1/\rho)<\frac{a}{2}$ \cite{SA}.
When $\epsilon_n$ does not grow in Eq. (\ref{ep2}), equation (\ref{de}) gives the condition on $K$ as $K>0$.

For PFF, the analysis is not straightforward;
we start with Eq. (\ref{deg}). By ignoring the coupling between $\epsilon_n$ and $\delta_n$ as well as
the fluctuations in the phase difference (i.e., $\Delta\phi^{(0)}_n\simeq\Delta\phi_{0}$), we obtain
\begin{equation}
\dot{\delta}_n = K \cos{(\Delta\phi_0)} \Delta\delta_n.
\end{equation} 
From this equation, we can find the stability condition, $|\Delta\phi_0|<\frac{\pi}{2}$.
Additionally, since the system is periodic ($\phi_{N+1}=\phi_1$), 
the condition $N \Delta\phi_0=2m\pi\quad(m=\pm1,\pm 2,\cdots)$ must be satisfied.
To summarize, stable $\Delta\phi_0$ can take the discrete values of 
$\Delta\phi_0=0,\pm\frac{2\pi}{N},\pm\frac{4\pi}{N},\cdots,\pm\left(\frac{\lfloor N/2\rfloor-2}{N}\right)\pi$,
where $\lfloor Y\rfloor$ is the floor function giving the maximum integer not larger than $Y$.
These conditions were verified by numerical simulations.
On the other hand, the coupling behavior of $\epsilon_n$ and $\delta_n$ is determined by the final term in Eq. (\ref{epg}).
If the time integral $\int_0^{T'}{\sin{\phi^{(0)}_n}dt}\neq 0$ 
(i.e., if the sine function is ``distorted"), the system can destabilize.
This distortion is also observed as a change of flux in the high-density regime (Fig. \ref{flowchange}. See also approximate analyses in Appendix \ref{sec:dis}.).
From these facts, we conclude that the high-density regime leads to coupling in the stability equation,
which may result in  DJF. 

\begin{figure}[t]
   \includegraphics[width=90mm]{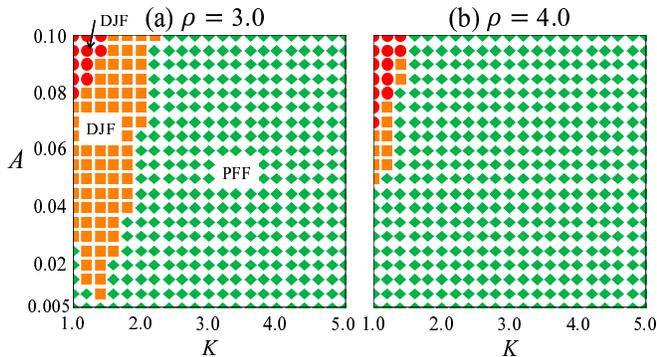}
 \caption{Phase diagram in the high-density regime.
 Initial conditions are given by the expression of PFF with $\Delta\phi_0 = \frac{2\pi}{100}$, ($x_{n,1},\phi_{n,1}$).
If we take SFF as initial conditions ($\Delta\phi_0=0$), the flow is always stable.
  Green diamonds indicate the parameter region where PFF is stable. 
Orange squares and red circles correspond to the emergence of DJF, and the latter include collisions of walkers. 
The same diagrams for low- and middle-density regimes only show PFF and OJF, respectively.
}
 \label{DJFP}
\end{figure}

In Fig. \ref{DJFP} we illustrate the phase diagrams of the system by simulations, using the expression for PFF $(x_{n,1},\phi_{n,1})$ 
(whose derivation is found in Appendix \ref{sec:ap}) as initial conditions,
to see whether the flow is stable or not:
\begin{eqnarray}
x_{n,1}&=&v_0t+\frac{n}{\rho} + \frac{Aa}{\dot{\phi}_0\sqrt{a^2+\dot{\phi}_0^2}}\sin{\Psi(t)}\nonumber\\
&&+\frac{2Aa^2V'(\frac{1}{\rho})\sin{(\frac{\Delta\phi_0}{2})}}{\dot{\phi}_0^2(a^2+\dot{\phi}_0^2)}\sin{(\Psi(t)-\varphi_0+\frac{\Delta\phi_0}{2})},\nonumber\\
\phi_{n,1}&=&\dot{\phi}_0t + \Delta\phi_0n\nonumber\\
&&+\frac{2Aa\omega'(\frac{1}{\rho})\sin{\frac{\Delta\phi_0}{2}}}{\dot{\phi}_0^2\sqrt{a^2+\dot{\phi}_0^2}}\sin(\Psi(t)+\frac{\Delta\phi_0}{2}),\nonumber
\end{eqnarray}
where $v_0=V(1/\rho)+A,\dot{\phi}_0 = \omega(1/\rho)+K\sin{\Delta\phi_0},\varphi_0=\tan^{-1}(\dot{\phi}_0/a)$, and $\Psi(t)=\dot{\phi}_0t-\varphi_0+\Delta\phi_0n$.
It is found that the system destabilizes for small $K$ and large $A$, and for especially large values of $A$, particles even collide with each other in DJF.
Thus we can conclude that the increase of $A$ and $K$ are connected to the destabilization and stabilization of the system, respectively.
In addition, increase of particle density reduces the DJF regime,
because it decreases space for ``over acceleration" of particles that is indispensable for DJF.


\section{conclusive discussion}\label{con}
In summary, we have proposed a simple model to show that the restriction coming from locomotion largely affects 
the collective behavior of SPPs. 
The model has various types of flow, including a novel type of complex flow called disordered jam flow (DJF).
Disordered jam flow appears when the particle density is high; this flow is caused by 
 coupling between velocities and phases of particles.
We also find that particle flux is strongly affected by phase anchoring.
On the other hand, some issues remain to be clarified in future studies.
These include, synchronization phenomena on inhomogeneous distributions of 
intrinsic optimal velocity and angular velocity, quantitative analysis for phase anchoring and its effect
on flux, etc. 
These points should be addressed in detail in future works.

Although pedestrian locomotion in a crowd is not yet fully understood,
a real pedestrian may adapt the crowd velocity and step length by considering the environment of other pedestrians (not only the nearest predecessor).
Furthermore, the environment may affect a pedestrian through a strong psychological repulsive force \cite{Prox}.
We believe that by including the effects of phase, models for pedestrians and other animals can be improved. 
Moreover, the synchronization of pedestrians through a bridge \cite{Mille,Mille2} might be an interesting problem.
It is also attractive to extend this model to two dimensions using the two-dimensional OV model \cite{POV},
which generalization would require additional rules of phase synchronization.

\section{acknowledgement}
We thank Hiroshi Kori for his insightful comments on this study.

\appendix
\begin{widetext}
\section{Approximate description of PFF}\label{sec:ap}
In this section we give equations of a particle trajectory in PFF approximately.
The trajectory satisfies the following system equations:
\begin{eqnarray}
\ddot{x}_{n}&=& a\{ V(\Delta x_{n}) + A (\cos{\phi_n}+1) - \dot{x}_n\}\label{se1s},\\
\dot{\phi}_{n} &=& \omega(\Delta x_{n}) + K \sin{\Delta\phi_{n}}.\label{se2s}
\end{eqnarray}
First, we assume that the deviation of each $\Delta x_{n}$ is small compared to its absolute value, and that 
the phase difference $\Delta \phi_n$ is constant ($\Delta \phi_0$).
In this assumption, $\Delta x_{n} = \frac{1}{\rho},\dot{\phi}_{n} = \omega(\frac{1}{\rho}) + K \sin{\Delta\phi_0}\equiv \dot{\phi}_0$(const.).
Then, the system equations ($\ref{se1s}$) and ($\ref{se2s}$) can be easily integrated:
\begin{eqnarray}
\phi_{n,0} &=& \dot{\phi}_0t + \Delta \phi_0 n,\\
x_{n,0}(t) &=& \frac{Aa}{\dot{\phi}_0\sqrt{a^2 + \dot{\phi}^2_0}}\sin{(\dot{\phi}_0t-\varphi_0+\Delta \phi_0 n)}+v_0t + \frac{n}{\rho},\\
\dot{x}_{n,0}(t) &=& \frac{Aa}{\sqrt{a^2+\dot{\phi}^2_0}}\cos{(\dot{\phi}_0t-\varphi_0+\Delta \phi_0 n)}+ v_0.
\end{eqnarray}
Here we put $v_0\equiv V(\frac{1}{\rho}) + A$ and $\varphi_0 = \tan^{-1}({\frac{\dot{\phi}_0}{a}})$.
These equations give exact trajectories of  particles when $\Delta \phi_0 =0$ (SFF). However these expressions are not enough for understanding
the dynamics of PFF in high-density cases. We next give a more accurate approximation based on these equations; 
the deviation of particle distance $\Delta x_n$ is taken into consideration, while the deviation of phase-difference is still ignored.
Using the previous result we approximately give $\Delta x_n = x_{n+1,0}-x_{n,0}$ in advance:
\begin{eqnarray}
\Delta x_n-\frac{1}{\rho} &\simeq&\frac{Aa}{\dot{\phi}_0\sqrt{a^2 + \dot{\phi}^2_0}}\left[\sin{(\dot{\phi}_0t-\varphi_0+\Delta \phi_0 n+\Delta \phi_0)}\right.
\left.-\sin{(\dot{\phi}_0t-\varphi_0+\Delta \phi_0 n)} \right]\\
&=&\frac{2Aa\sin{\frac{\Delta\phi_0}{2}}}{\dot{\phi}_0\sqrt{a^2 + \dot{\phi}^2_0}}\cos{\left(\dot{\phi}_0t-\varphi_0+\Delta \phi_0 n+\frac{\Delta\phi_0}{2}\right)}\\
&=&A_{\Delta}\cos{\left(\dot{\phi}_0t-\varphi_0+\Delta \phi_0 n+\frac{\Delta\phi_0}{2}\right)},\label{dxf}
\end{eqnarray}
where $A_{\Delta}$ is the amplitude of oscillation of $\Delta x_n$.
Then, the equation for $\phi_n$ ($\ref{se2s}$) is now rewritten as
\begin{eqnarray}
\dot{\phi}_{n,1} &=& \omega(1/\rho) + K\sin{\Delta \phi_0}+ \omega '(1/\rho)A_{\Delta}\cos{\left(\dot{\phi}_0t-\varphi_0+\Delta \phi_0 n+\frac{\Delta\phi_0}{2}\right)},\label{phif}
\end{eqnarray}
which can be solved as:
\begin{equation}
\phi_{n,1} = \phi_{n,0} + \frac{2Aa\omega '(1/\rho)\sin\frac{\Delta\phi_0}{2}}{\dot{\phi}_0^2\sqrt{a^2+\dot{\phi}_0^2}}\sin{\left(\dot{\phi}_0t-\varphi_0+\Delta \phi_0 n+\frac{\Delta\phi_0}{2}\right)}.
\end{equation}
Here we assumed $\Delta x_n - \frac{1}{\rho} \ll1$. 
In the same manner, we can derive  a more accurate expression for $\Delta x_n$:
\begin{equation}
x_{n,1} = x_{n,0}+\frac{2Aa^2V'(1/\rho)\sin(\frac{\Delta\phi_0}{2})}{\dot{\phi}_0^2(a^2+\dot{\phi}_0^2)}\sin(\dot{\phi}_0 t-2\varphi_0 + \Delta\phi_0 n +\frac{\Delta \phi_0}{2}).
\end{equation}
Note that the effect of these correction terms increases as $|\dot{\phi_0}|$ (or $\Delta \phi_0$) becomes small (large), 
namely, the first approximations presented in this section do not agree with actual trajectories.
As shown in Fig. \ref{Ad}, the approximation is valid for large $K$ and small positive $\Delta\phi_0$.

We improved the approximation using $(x_{n,0},\phi_{n,0})$ to obtain $(x_{n,1},\phi_{n,1})$.
In the same procedure, further improvement can be made by calculating the series of trigonometric
functions $(x_{n,k+1},\phi_{n,k+1})$ from $(x_{n,k},\phi_{n,k})$ if their amplitude is small enough to
allow the Taylor expansion of the optimal-velocity functions, while it is far more difficult to describe the system where this assumption is not valid.

In the main text, we used $x_{n,1}$ and $\phi_{n,1}$ as initial conditions to plot the phase diagrams.

\begin{figure*}[tb]
   \includegraphics[width=150mm]{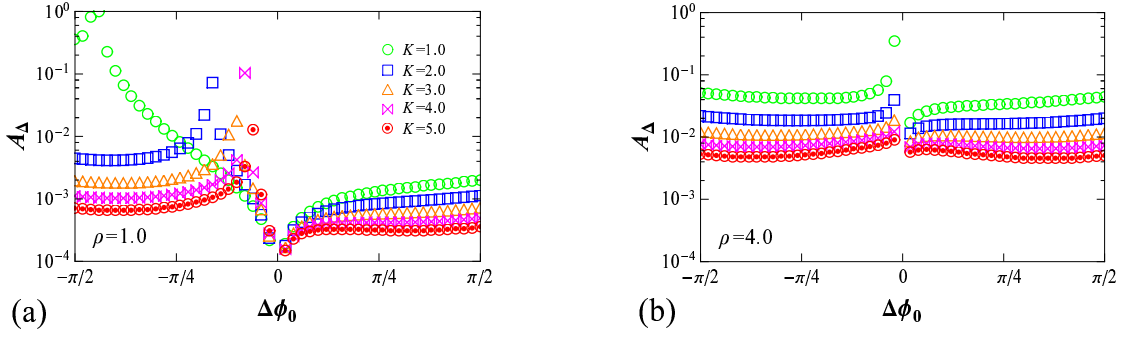}
 \caption{Relationship between $A_{\Delta}$ and $\Delta \phi_0$ for the cases of $A=0.05$.}
 \label{Ad}
\end{figure*}

\begin{figure*}[tb]
   \includegraphics[width=150mm]{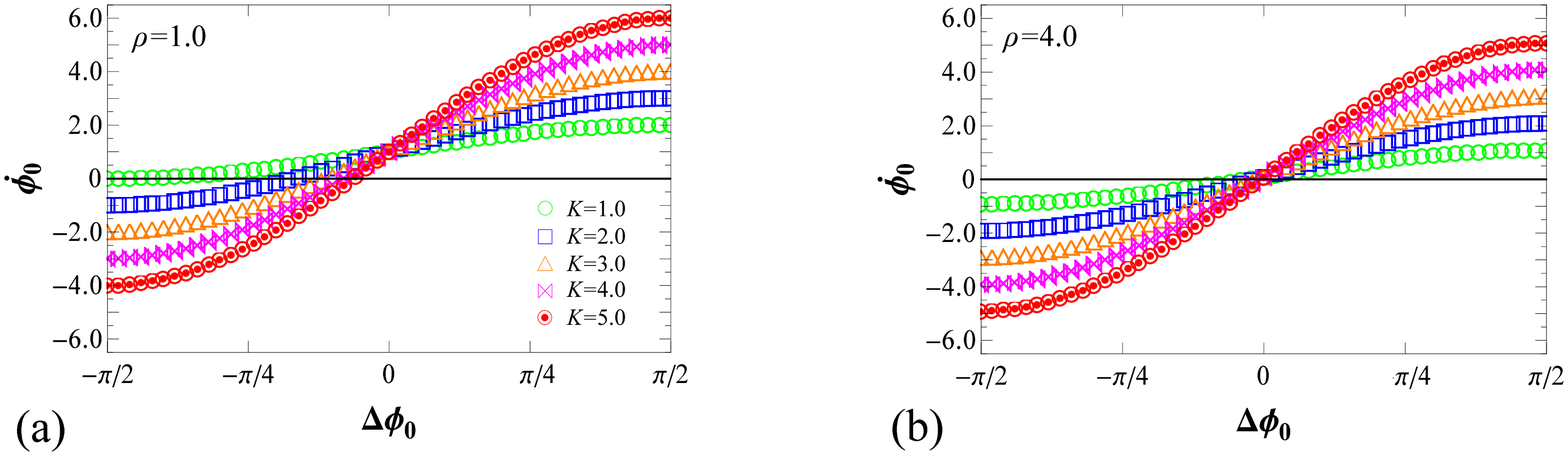}
 \caption{Relationship between $\dot{\phi}_0$ and $\Delta\phi_0$ for the cases of $A=0.05$.}
 \label{dphi}
\end{figure*}

\section{Distortion of trajectory and its effect on particle flux}\label{sec:dis}
Average velocity of a particle in a steady flow is given by integrating Eq. (\ref{se1s}) over a period of one ``step", $\mathcal{T}$:
\begin{equation}
\frac{1}{\mathcal{T}} \int_\mathcal{T}{\dot{x}_ndt}=\frac{1}{\mathcal{T}}\int_\mathcal{T}{V(\Delta x_n)dt}+A+\frac{1}{\mathcal{T}}\int_\mathcal{T}{A \cos{\phi_n}}dt.
\end{equation}
By using the first-order approximation ($x_{n,0},\phi_{n,0}$) and assuming that the amplitudes of $\Delta x_n$ and $A_{\Delta}$ are small enough, 
we can estimate the average velocity as
\begin{equation}
\frac{1}{\mathcal{T}} \int_\mathcal{T}{\dot{x}_ndt}\simeq v_0 + \frac{1}{T}\int_\mathcal{T}{A \cos{\phi_{n,1}}}dt.
\end{equation}
Therefore the particle flux in PFF is affected by the fluctuation of $\dot{\phi}_n$ from $\phi_{n,0}$
, by which the final integral in the above equation remains nonzero.
Now the degree of the distortion is characterized by Eq. (\ref{phif}).
For example, we see the difference between the functions $\cos{\phi_{n,0}}$ and $\cos{\phi_{n,1}}$ 
when the focal phase difference $\Delta \phi_0$ is a small positive number that leads to small $\dot{\phi}_0$.
The former function is at its maximum at $\dot{\phi}_0t+\Delta\phi_0 n = 2 m\pi$, at which the 
phase velocity of  the modified approximation $\phi_{n,0}$ is faster than $\dot{\phi}$:
\begin{equation}
\dot{\phi}_{n,1}|_{\phi_{n,0}=2m\pi} = \dot{\phi}_{0} + \frac{2Aa\omega'(1/\rho)\sin{\frac{\Delta\phi_0}{2}}}{\dot{\phi}_0\sqrt{a^2 + \dot{\phi}^2_0}} \cos{\left( \frac{\Delta\phi_0}{2}-\varphi_0\right)}>\dot{\phi_0}.
\end{equation}
The contribution of the phase difference, $\frac{\Delta\phi_0}{2}-\varphi_0$, is depicted in Fig. \ref{cos}.
In the same manner, we can find that when $\phi_0 = (2m+1)\pi$, $\dot{\phi}_{n,1}$ is slower than $\dot{\phi}$.
Consequently, the positive part of the cosine function (that corresponds to the approximation $\phi_{n,0}$) shrinks while its negative part is amplified (see Fig. \ref{dist}),
resulting in a smaller value of the integral, and the  particle flux decreases as shown in Fig. 3 in 
the main text. 
\begin{figure}[tb]
   \includegraphics[width=90mm]{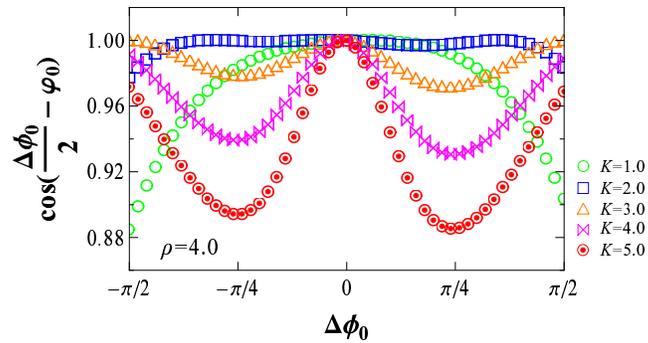}
 \caption{$\cos{\left(\frac{\Delta\phi_0}{2}-\varphi_0\right)}$ vs $\Delta \phi_0$ for the cases of $\rho=4.0$ and $A=0.05$.}
 \label{cos}
\end{figure}
\begin{figure}[tb]
   \includegraphics[width=60mm]{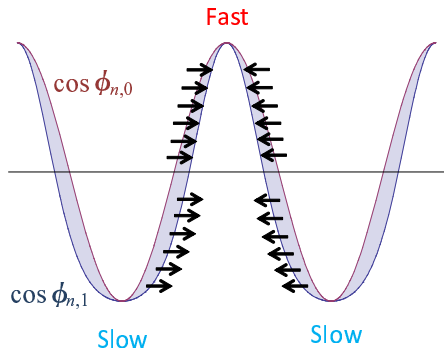}
 \caption{Effect of fluctuation of $\dot{\phi}_n$ on the distortion of the cosine function.}
 \label{dist}
\end{figure}
Thus the decrease in the flux around $\Delta \phi_0 = 0$ can be explained in terms of 
the interaction between oscillations. 
However, this theory cannot account for the cases of $J/J_s>1$, for which we have to take large deformations
of trajectories into account.
\end{widetext}

\bibliography{basename of .bib file}

\end{document}